\documentclass[showpacs,preprintnumbers,amsmath,amssymb,nofootinbib]{revtex4}

\usepackage{graphicx}
\usepackage{dcolumn}
\usepackage{bm}
\usepackage{latexsym}
\usepackage{epsfig}
\usepackage{pbox}
\usepackage{multirow}
\usepackage{rotating}
\usepackage{bm}

\newcommand{\be}{\begin{equation}}
\newcommand{\ee}{\end{equation}}
\newcommand{\beqq}{\setlength\arraycolsep{2pt}\begin{eqnarray}}
\newcommand{\eeqq}{\vspace{0cm} \end{eqnarray}}
\newcommand{\bea}{\begin{eqnarray}}
\newcommand{\eea}{\end{eqnarray}}

\newcommand{\lambdab}{\stackrel{\neg}{\lambda}}
\newcommand{\xib}{\stackrel{\neg}{\xi}}
                
\newcommand{\xb}{\bar{x}}
\newcommand{\vb}{\bar{v}}
\newcommand{\yb}{\bar{y}}

\begin{document}

\title{A new approach on the stability analysis in ELKO cosmology}

\author{A. Pinho S. S.} \email{alexandre.pinho510@gmail.com}
\author{S. H. Pereira} \email{shpereira@gmail.com}
\affiliation{Faculdade de Engenharia de Guaratinguet\'a \\ UNESP - Univ. Estadual Paulista ``J\'ulio de Mesquita Filho''\\ Departamento de F\'isica e Qu\'imica\\ Av. Dr. Ariberto Pereira da Cunha 333 - Pedregulho\\
12516-410 -- Guaratinguet\'a, SP, Brazil}
\author{J. F. Jesus} \email{jfjesus@itapeva.unesp.br}
\affiliation{Universidade Estadual Paulista ``J\'ulio de Mesquita Filho'' -- Campus Itapeva \\
Rua Geraldo Alckmin 519, 18409-010, Vila N. Sra. de F\'atima, Itapeva, SP, Brazil}



\begin{abstract}
In this work it has been developed a new approach to study the stability of a system composed by an ELKO field interacting with dark matter, which could give some contribution in order to alleviate the cosmic coincidence problem. It is assumed that the potential which characterizes the ELKO field is not specified, but it is related to a constant parameter $\delta$. The strength of the interaction between matter and ELKO field is characterized by a constant parameter $\beta$ and it is also assumed that both ELKO field as matter energy density are related to their pressures by equations of state parameters $\omega_\phi$ and $\omega_m$, respectively. The system of equations is analysed by a dynamical system approach. It has been found the conditions of stability between the parameters $\delta$ and $\beta$ in order to have stable fixed points for the system for different values of the equation of state parameters $\omega_\phi$ and $\omega_m$, and the results are presented in form of tables. The possibility of decay of ELKO field into dark matter or vice versa can be read directly from the tables, since the parameters $\delta$ and $\beta$ satisfy some inequalities. It allows us to constrain the potential assuming that we have a stable system for different interactions terms between the ELKO field and dark matter. The cosmic coincidence problem can be alleviated for some specific relations between the parameters of the model.

\end{abstract}

\maketitle


\section{Introduction}

Similarly to many segments in science, there are still open questions in modern cosmology to be answered. One of the greatest puzzles concerns the composition of the universe, which can be roughly divided into three components, namely the ordinary baryonic matter ($\approx 5\%$), the Dark Matter (DM) ($\approx 25\%$) and the Dark Energy (DE) ($\approx 70\%$), according to the most accepted models \cite{SN,reviewDE,reviewDM}. Currently, we have direct access only to the former component, but there are many attempts to detect DM particles, since it behaves exactly like the usual baryonic matter, although not interacting electromagnetically \cite{Taoso,bookDM}. DE is something even more mysterious, since their behaviour is gravitationally repulsive \cite{DER,DEV}. From a theoretical point of view it is very common to interpret these entities as being some kind of field (scalar fields \cite{Carvalho,scalar,matos} or spinor fields \cite{DEspinor}, for example). Other models also consider the possibility of interaction between DM and DE \cite{das,Feng,DM,DE1,DE2,DE3,DE4,abwa,guo,Ioav,saulofernando}, which could alleviate the coincidence problem for instance. 

Besides the usual scalar fields, some recent works have shown that there are some classes of Non-Standard Spinors with interesting properties which could be useful in order to describe both DM as DE. One of such spinors is called ELKO, from German \textit{\textbf{E}igenspinoren des \textbf{L}adungs\textbf{k}onjugations\textbf{o}perators}, which has the property to be an eigenspinor of change conjugation and parity, possessing non-locality \cite{AHL1,AHL2,AHL3,BOE1,BOE2,BOE5,BOE3,BOE4,FABBRI,BOE6,BOE7,GREDAT,BASAK,roldao1,roldao2,roldao3,julio1,js}. It satisfies ${ \left( CPT \right)  }^{ 2 }=-I $ and also has some other properties, as a spin one half and a mass dimension 1, which makes it a good candidate to a particle with small probability of interacting with Standard Model particles, exactly as desired for the DM particles and maybe also to DE. The searching for ELKO dark matter at the CERN LHC has also been addressed recently \cite{dias1,alves1}. 

The ELKO spinor field, although being a fermionic field, can be factored out in a time dependent scalar field multiplied by a four components spinor field. Its scalar part drives the evolution of the pressure and energy density associated to the ELKO field, and it is through these physical quantities that ELKO contributes to cosmological equations. Here we will consider the ELKO field as a candidate to DE interacting with a DM field. The recent concern in this kind of exotic non-standard field as a good candidate to DE or DM is due to the fact that the scalar part of ELKO spinors has a much richer structure than the standard scalar field when applied to cosmology \cite{Carvalho,scalar,matos}. The coupled system of equations involving the interaction of DE and DM plus the Friedmann equations are much involved, even for a standard scalar field \cite{matos}. Such system is much more involved when dealing with ELKO fields. Due to its complexity, dynamical system analysis have been applied in order to deal with ELKO field as a possible candidate to DM or DE. Although exact solutions are not found even within this method for most cases, obtaining stability points that turns the system well behaved provides useful information about some undetermined parameters of the system. In our specific case we are interested in some conditions to be satisfied by the potential of the ELKO spinor field in order to maintain the system stable. Maintaining the system stable around some fixed points indicates that the involved system of equations has stable solutions for that specific region where the variables are evolving. In our case it can be interpreted as possible solutions where the ELKO field and DM can coexist, indicating a solution to the cosmic coincidence problem around these fixed points.

An interesting aspect concerning the ELKO field dynamics is that the choice of variables is an important question when one is looking for stable points of the dynamical system. Such a search for adequate variables depends in general on the structure of the physical quantities involved, like the energy density, pressure and Friedmann constraints. In some recent works \cite{basak,WEI,sadja}, different choices of variables for interacting systems concerning the ELKO field have shown that there are no stable points in order to explain the cosmic coincidence problem. In \cite{saj} it was proposed a new method of analysis based on a constant parameter that leads to stable points under some conditions. In the present work the same method of \cite{saj} has been applied for a new set of dynamic variables and stable fixed points have been found for the system, including new interaction terms. This opens the possibility to alleviate the cosmological coincidence problem considering the ELKO field interacting with DM if the conditions found here are satisfied.


The new set of dynamical variables for the ELKO field used in this work is independent of the choice of potential. The potential dependence is set to a constant parameter $\delta$. Besides that it is assumed that the pressure and energy density of the ELKO field satisfies an equation of state of the form $p_\phi = \omega_\phi \rho_\phi$, where $p_\phi$ and $\rho_\phi$ are the ELKO field pressure and energy density, respectively. The dark matter content is described by an energy density $\rho_m$ that satisfies an equation of state of the type $p_m = \omega_m \rho_m$. It is also assumed an interaction between ELKO field and DM, characterized by a constant parameter $\beta$. The conditions for stability between $\delta$ and $\beta$ have been studied, depending on the type (radiation, dust, vacuum or ultra-relativistic matter) of the thermodynamic equation of state parameters $\omega_m$ and $\omega_\phi$. 


Let us finish citing that the most general ELKO theory applied to cosmology should include torsion \cite{BOE2,FABBRI,FABBRI1}. This generalization is possible constructing an action where connections are not more symmetric at all. Such antisymmetric part of connection is defined as the torsion tensor, which is a rotation of a vector when it is parallel transported. Nevertheless problems with torsion have not been properly studied yet, although in some works we can see that torsion could play a cosmological constant role, accelerating the universe. It is not also well known if torsional effects appear to affect the dynamics only at the Planck scale. 

The paper is organized as follows. In Section II we start presenting a general scenery including torsion and then we restrict to the torsionless case with the basic equations for a two fluid model, namely DM and ELKO field equations related to our cosmological applications in a spatially flat Friedmann-Lema\^itre-Robertson-Walker (FLRW) background. We present the pressure and energy density expressions, as well as the Friedmann equations and the conservation equations including the interaction between dark matter and ELKO field. We also define the variables concerning the dynamical system equations to be analysed. Section III contains the main results. We study the stability of the dynamical system by imposing the restriction that the potential of the ELKO field is related to a constant parameter $\delta$. This analysis allows us to study the conditions to be satisfied by $\delta$ and $\beta$ for each equation of state parameters of DM and ELKO field, in order to maintain stability. For each kind of interaction we present the results in form of tables, specifying the equation of state parameter and conditions of stability to be satisfied. For DM and ELKO field we restrict the equation of state parameters to vacuum, dust, radiation and ultra-relativistic matter. In Section IV we conclude with some remarks.

\section{ELKO cosmology as a dynamical system}
We begin introducing the general Einstein-Cartan action:
\begin{equation}
S = \int \left( \frac{1}{2\kappa^2}\tilde{R} + \tilde{\mathcal{L}}_{mat} \right)\sqrt{-g}d^4 x \,,
\label{action}
\end{equation}
where tilde denotes the presence of torsion terms into covariant derivatives. We have also that $\kappa^2\equiv8\pi G$ with the normalization $c=1$. The matter source (DM and ELKO) is present in the matter Lagrangian, $\tilde{\mathcal{L}}_{mat}$. In this work we have divided such Lagrangian into two parts. One part concerns the DM (including baryonic matter), $\tilde{\mathcal{L}}_{DM}$, with an attractive behaviour. The other part is playing the role of a DE fluid, here represented by the ELKO spinor Lagrangian $\tilde{\mathcal{L}}_{\lambda}$: 
\begin{equation} 
\tilde{\mathcal{L}}_\lambda=\frac{1}{2}g^{\mu\nu}\tilde{\nabla}_{(\mu}{\lambdab} \tilde{\nabla}_{\nu)} \lambda - V({\lambdab}\lambda).
\label{tilde_L}
\end{equation}

The covariant derivative acting on a spinor in presence of torsion is
\begin{equation}
\tilde{\nabla}_{a}\lambda = \partial_{a}\lambda - \frac{1}{4}\Gamma_{a}\lambda + \frac{1}{4}K_{abc}\gamma^b \gamma^c \lambda
\end{equation}
where $\Gamma_a=\frac{1}{4}\Gamma_{abc}[\gamma^b\gamma^c,\,\gamma^b\gamma^c]$ and $K_{abc}$ is known as contortion tensor, which represents the antisymmetric part of the Christoffel symbol. Then it is possible to find an analogous of Einstein equation by doing variation of action (\ref{action}):
\begin{equation}
\tilde{G}_{ij}=\tilde{R}_{ij}-\frac{1}{2}\tilde{R}g_{ij}=\kappa^2\Sigma_{ij},
\label{E_Tensor_tor}
\end{equation}
and also an equation for the spin angular momentum tensor $\tau^{ij}_k$,
 \begin{equation}
{{T}^{ij}}_{k}+\delta^i_k {T}^{j\;l}_{\,\;l}-\delta^j_k {T}^{i\;l}_{\,\;l}=\frac{{\tau^{ij}}_{k}}{\kappa^2}\,,
\end{equation}
obtained by variation of the Einstein-Cartan action with respect to the contortion tensor ${{K}^{ij}}_{k}$, which is related to the torsion by ${T_{\mu\nu}}^{\lambda}={1\over 2}({{K}_{\nu\mu}}^{\lambda}-{{K}_{\mu\nu}}^{\lambda})$.

As well as in the ordinary Einstein equation, the right hand side of (\ref{E_Tensor_tor}) represents the matter source, however with a new form due to torsion. It is given by
\begin{equation} 
\Sigma_{ij} = \sigma_{ij}+\left(\tilde{\nabla}_{k}-{K_{lk}^l} \right)\left({\tau_{ij}}^{k} + {{{\tau}_{j}}^{k}}_{i} + {{\tau}^{k}}_{ij} \right),
\end{equation}
where $\sigma_{ij}$ corresponds to the energy momentum tensor related to $\tilde{\mathcal{L}}_{mat}$. 

In a flat FLRW metric, $ds^2=dt^2-a(t)^2(dx^2+dy^2+dz^2)$, where $a(t)$ is the scale factor, the Christoffel symbols are $\Gamma^i_{ti}=\dot{a}/a$ and $\Gamma^t_{ii}=\dot{a}a$, where $i=x,\,y,\,z$, and
\begin{equation}
{\Sigma}_{tt}=\frac{\tilde{\rho}}{2}\,\,, \hspace{1cm} {\Sigma}_{ii}=\frac{\tilde{p}}{2 a^2},
\end{equation}
where $\tilde{\rho}$ and $\tilde{p}$ are the energy density and the pressure of the matter sources, in our case composed by DM and ELKO field, which can be approximated as a perfect fluid on large scales. If the ELKO field density depends only on cosmological time, as it must obey cosmological principle, the allowed components of the torsion tensor are reduced to
\begin{equation}
T_{110}=T_{220}=T_{330}=h(t),
\end{equation}
\begin{equation}
T_{123}=T_{312}=T_{231}=f(t),
\end{equation}
where $h$ and $f$ are functions carrying the torsion contributions. Finally, the Friedmann equations with torsion are given by
\begin{equation}
H^2 + 2hH + h^2 - f^2 = \frac{\kappa^2}{3}\tilde{\rho}\,,\label{T1}
\end{equation}
\begin{equation}
\dot{H} - hH + \dot{h} - h^2 + f^2 = -\frac{\kappa^2}{2}\left(\tilde{\rho} +\tilde{p}\right)\,,\label{T2}
\end{equation}
with $H=\dot{a}/a$.

From the above general treatment, we see that even assuming homogeneity and isotropy, the torsion functions $h(t)$ and $f(t)$ yet carry non trivial contributions to the time component of the evolution equations. Thus, even for an ELKO field that behaves dynamically as scalar, its contributions should be taken into account. When it is used into cosmology, it is important to emphasize that they are spinors, nevertheless with important consequences: the most important of them is the fact that a totally-coupled ELKO field must have torsion, and with torsion additional terms are present in the Friedmann equations, as seen above.

However, in order to study the possibility of ELKO field interaction with DM, intending to alleviate the cosmic coincidence problem, we will restrict our study to the torsionless case. It is evident that the presence of torsion should lead to more general results.

When the torsion terms are dropped out, it is easy to see that we obtain the ordinary Friedmann equation. Before doing this, let us just make a simplification on the ELKO field. As it has been done in recent works \cite{BOE3,BOE4,BOE6,BOE7,WEI}, we will restrict the ELKO spinor field to the form $\lambda\equiv \phi(t)\xi$ and $\lambdab\equiv \phi(t)\xib$, where $\xi$ and $\xib$ are constant spinors\footnote{In \cite{js} it has been presented exact solutions to ELKO spinor in spatially flat FLRW expanding spacetimes, and it has been shown that such factorisation of the ELKO field time component is possible for some types of scale factors.}. Then, in the torsion-free case, we have $h\to 0$, $f\to 0$, $\tilde{\rho}\to \rho_m + \rho_\phi$ and $\tilde{p}\to p_m + p_\phi$, where the subscripts $_m$ and $_\phi$ indicate, respectively, the DM and ELKO field contributions to energy density and pressure.

Thus, the FLRW equations in a spatially flat background without torsion are given by:
\be
H^2 = \frac{\kappa^2}{3}\left(\rho_m +\rho_{\phi}\right)\,,\label{H2}
\ee
\be
\dot{H}=-{\kappa^2 \over 2}(\rho_m +p_m +\rho_{\phi}+p_\phi)\,.\label{Hdot} 
\ee
We will assume that the DM and ELKO field satisfy equations of state of the form $p_m=\omega_m \rho_m$ and $p_\phi=\omega_\phi \rho_\phi$, respectively.
Due to the homogeneity of the field ($\partial_i \phi =0$), the equation of motion that follows from (\ref{tilde_L}) is substantially simplified to
\be
\ddot{\phi}+3H\dot{\phi}-{3\over 4}H^2\phi+V_{,\phi}=0\,,\label{eqphi}
\ee
where $V_{,\phi}\equiv dV/d\phi$. The pressure and energy density of the ELKO field are given by \cite{BOE6}:
\be
p_\phi={1\over 2}\dot{\phi}^2-V(\phi)-{3\over 8}H^2\phi^2-{1\over 4}\dot{H}\phi^2-{1\over 2}H\phi\dot{\phi}\,,\label{press}
\ee
\be
\rho_\phi={1\over 2}\dot{\phi}^2+V(\phi)+{3\over 8}H^2\phi^2\, .\label{rho}
\ee

The continuity equations for DM and ELKO field are:
\beqq
\dot{\rho}_m+3H(1+\omega_m)\rho_m=Q\,,\label{rhom}
\eeqq
\beqq
\dot{\rho}_\phi+3H(1+\omega_\phi)\rho_\phi=-Q\,,\label{rhophi}
\eeqq
where $Q$ stands for a possible interaction term between the DM and ELKO field. If $Q=0$, there is no interaction and both components evolve separately. If $Q>0$, the ELKO field decays into DM, an interesting scenery at the inflation, and if $Q<0$, DM decays into ELKO field (or DE).  

Exact solutions to the system of coupled equations (\ref{eqphi})-(\ref{rhophi}) are not available. For this reason we make a stability analysis based on dynamical system approach.

We define new variables:
\be
x={\kappa \dot{\phi}\over \sqrt{6}H}\,, \hspace{1cm} y={\kappa \sqrt{\tilde{V}}\over \sqrt{3}H}\,, \hspace{1cm} v={\kappa \sqrt{\rho_m}\over \sqrt{3}H}\,,\label{xyv}
\ee
where we have redefined the potential as $\tilde{V}=V+{3\over 8}H^2\phi^2$. The Friedmann equation (\ref{H2}) can be written as a constraint equation
\be
x^2+y^2+v^2 = 1\,,\label{constr}
\ee
or in terms of the densities parameters, $\Omega_\phi + \Omega_m =1$, where
\be
\Omega_\phi = {\kappa^2 \rho_\phi\over 3 H^2}= x^2 + y^2\,, \hspace{1cm}\Omega_m = {\kappa^2 \rho_m\over 3 H^2}=v^2 \,.\label{Omega}
\ee

The equations (\ref{Hdot}), (\ref{rhom}) and (\ref{rhophi}) can be written as a dynamical system of the form:
\begin{equation}
x'=\frac { 3 }{ 2 }(\omega_m - \omega_\phi)v^2 x-\left[ \frac { 3 }{ 2 } \left( 1+\omega_\phi  \right) +\frac { \lambda  }{ 2H }  \right] \frac { { y }^{ 2 } }{ x } -\frac { { Q }_{ 1 } }{ x } 
\label{xline}
\end{equation}

\begin{equation}
v'=\frac { 3}{ 2 }(\omega_\phi-\omega_m)(1-v^2)v+\frac { { Q }_{ 1 } }{ v }\,,
\label{vline}
\end{equation}

\begin{equation}
y'=\left(  \frac { 3 }{ 2 } \left( 1+\omega_\phi  \right) +\frac { 3 }{ 2 } \left( \omega_m -\omega_\phi  \right) { v }^{ 2 }+\frac { \lambda  }{ 2H }  \right) y
\label{yline}
\end{equation}
where $\lambda = {\dot{\tilde{V}}\over \tilde{V}}$, $Q_1 = {\kappa^2 Q\over 6 H^3}$ and $'$ stands for the derivative with respect to $N\equiv \ln a$, such that $f'= \dot{f}/H$ for any function $f$.

It is easy to see that the simple change of variables (\ref{xyv}) should not be sufficient to rewrite the Friedmann equations (\ref{T1}) and (\ref{T2})  in a simple form with the torsion terms. It would be necessary at least two new variables concerning $f$ and $h$, and also the system of equations (\ref{xline})-(\ref{yline}) would be much more complicated in the torsion case.

\section{Stability Analysis}

The stability analysis of the above set of dynamical equations consists of finding out fixed points $\bar{x}$, $\bar{v}$ and $\bar{y}$  that makes $x'$, $v'$ and $y'$ equal to zero. In the last section we have defined a three dimensional system according to our variable choice. However, due to the Friedmann constraint (\ref{constr}) the system can be reduced to a two dimensional one. 

Before we proceed, let us examine carefully the dynamical system (\ref{xline})-(\ref{yline}). We see that, in addition to the dynamical variables $x$, $v$ and $y$ we also have the factor ${\lambda\over 2H} = {\tilde{V}'\over 2\tilde{V}}$. The presence of the $'$ derivative shows that such term is also a dynamical variable, which should also be taken into account. However, $\tilde{V}$ depends on the potential but potential is not specified in our analysis, thus we can not deal with this new variable. In order to avoid this problem, we set the additional assumption related to the potential, $-\frac { \lambda  }{ 2H } \equiv \delta$, where $\delta$ is a constant parameter. Such imposition just reflects our ignorance on the potential $V(\phi)$.

Now it is easy to see that the resulting dynamical system, for a given interaction $Q_1(x,v,y)$, is written in terms of the dynamical variables $x$, $v$, $y$ and the constants $\omega_m$, $\omega_\phi$ and $\delta$. In order to analyse the stability of this system around fixed points $\xb$, $\vb$ and $\yb$ we must study the system satisfying $x'=0$, $v'=0$ and  $y'=0$. Notice that the equation (\ref{yline}) is independent of the interaction $Q_1$, thus the condition $y'=0$ can be achieved only if $\yb=0$ or $-{\lambda\over 2H} =\delta=\frac { 3 }{ 2 } \left( 1+\omega_\phi  \right) +\frac { 3 }{ 2 } \left( \omega_m -\omega_\phi  \right) { \vb }^{ 2 }$. The first condition can be satisfied only if $\tilde{V}=0$ (see (\ref{xyv})), but from the definition of $\lambda$ this leads to a divergent $\lambda$. So, we restrict ourselves to the second condition, namely
\be
\delta=\frac { 3 }{ 2 } \left( 1+\omega_\phi  \right) +\frac { 3 }{ 2 } \left( \omega_m -\omega_\phi  \right) { \vb }^{ 2 }\,,\label{delta}
\ee
where $y'=0$ even for $\yb\neq 0$. By imposing the above condition on the dynamical system (\ref{xline})-(\ref{yline}) we are left with a $2 \times 2$ system, since that $y' = 0$ is always satisfied:
\begin{equation}
x'= \left[\frac { 3 }{ 2 } (1+\omega_\phi)-\delta +\frac { 3 }{ 2 }(\omega_m-\omega_\phi)v^2  \right] x+\left[ \delta -\frac { 3 }{ 2 }(1+\omega_\phi) \right]{(1-v^2)\over x} -\frac { { Q }_{ 1 } }{ x }\, ,
\label{xline2}
\end{equation}

\begin{equation}
v'=\frac { 3}{ 2 }(\omega_\phi-\omega_m)(1-v^2)v+\frac { { Q }_{ 1 } }{ v }\,,
\label{vline2}
\end{equation}
where we have also used the Friedmann constraint (\ref{constr}).

In order to study such dynamical system it is worth to define its linearised matrix, with which one can determine the stability of a fixed point by just analysing its determinant and trace. Such mechanism is ensured by Hartmann-Grobman theorem \cite{booksystem}. Thus, in the neighbourhood of the fixed points we take infinitesimal displacements of variables from its fixed points,  $x\to \bar{x}+\delta x$ and $y\to \bar{v} + \delta v$, so that
\begin{eqnarray}
\left(\begin{array}{c}
\delta x'\\ \delta v'
\end{array}\right)=M
\left(\begin{array}{c}
\delta x\\ \delta v
\end{array}\right)\,
\label{delta_sys}
\end{eqnarray}
where $M$ is given by
\begin{eqnarray}
M=
\left(\begin{array}{cc}
a & \;\;b \\ 
c & \;\;d \\
\end{array}\right)
\label{M}
\end{eqnarray}
and

\begin{equation}
a=-\bigg[\delta -{3\over 2}(1+\omega_\phi)\bigg]{(1-\vb^2)\over \xb^2}+{Q_1\over \bar{x}^2} - {1\over \bar{x}}{\partial Q_1\over \partial\bar{x}}
\end{equation}

\begin{equation}
b={\partial x'\over \partial v}=3(\omega_m-\omega_\phi)\xb \vb + \bigg[3(1+\omega_\phi)-2\delta\bigg]{\vb\over \xb}-{1\over \bar{x}}{\partial Q_1\over \partial \bar{v}},
\end{equation}

\begin{equation}
c={\partial v'\over \partial x}={1\over \bar{v}}{\partial Q_1\over \partial \bar{x}},
\end{equation}

\begin{equation}
d={\partial v'\over \partial v}={3\over 2}(\omega_\phi-\omega_m)(1-3\vb^2)-{Q_1\over \bar{v}^2} + {1\over \bar{v}}{\partial Q_1\over \partial\bar{v}}.\label{d}
\end{equation}
 All variables carry a bar over them to show that the matrix $M$ is taken at the stable points which solve the system.

There is a simple way to know if the system described by the matrix $M$ is stable or not. It depends on the values of the determinant ($ \Delta $) and also on the trace ($ \tau $) of such matrix. When $\Delta > 0$, both eigenvalues have the same sign and if they are positive, the solution increases with the time evolution, indicating that the solutions diverge from the fixed point and consequently this point is classified as unstable. On the other hand, when both eigenvalues are negative, the solution goes to zero and the fixed point is stable. In order to know what kind of fixed point we are dealing with, it is necessary to check the value of the trace of matrix $M$. When $\tau>0$, it means that both eigenvalues are greater than zero, describing unstable points. However, when $\tau<0$ we have that they are negative and the point is stable. For the case where $\Delta <0$ we have that both eigenvalues have opposite signs and then the fixed point is in fact a saddle point. Finally, when $\Delta = 0$, at least one of the eigenvalues is zero and consequently nothing can be said about the stability of system.

Let us return to the dynamical system. Together with the equations (\ref{xline2}) and (\ref{vline2}), the assumption $-{\lambda\over 2H}=\delta$ leads to the new constraint for the fixed point $\vb$, according to (\ref{delta}): 
\begin{equation}
\bar { v } =\sqrt { \frac { 2\delta-3(1+\omega_\phi) }{3(\omega_m -\omega_\phi)  }  }.
\label{vbar}
\end{equation} 
But it is easy to see that such constraint already determines the value of the fixed point $\bar{v}$, since it depends only on the fixed parameters $\omega_m$, $\omega_\phi$ and $\delta$. Thus, in order to also satisfy the dynamical equation (\ref{vline2}), we have verified that this restriction tells us that $Q_1$, which represents the interaction, could not assume an arbitrary value, since that Eq. (\ref{vline2}) would not be always solved for an arbitrary $Q_1$. In other words, we have found that when $Q_1$ does not depend on $x$, not all fixed $\bar{v}$ that makes Eq. (\ref{vline2}) vanish also satisfy (\ref{vbar}), except for some very specific relations among the parameters $\omega_m$, $\omega_\phi$ and $\delta$. Such restriction on the interaction term $Q_1$ is not so strong, since the variable $x$ is proportional to $\dot{\phi}$, which characterizes the time variation of the field $\phi$, which is reasonable for an interacting theory.

In which follows, it will be analysed the stability conditions for different interaction terms between DM and ELKO field. The interaction terms are characterized by a dimensionless coupling constant $\beta$. We will search for stability conditions between the parameters $\delta$ and $\beta$ for different equation of state parameters $\omega_m$ and $\omega_\phi$. Besides stability conditions characterized by negative eigenvalues of the matrix of perturbation $M$, we impose the additional reality condition on the parameters (\ref{xyv}), namely we will impose $\xb^2>0$, $\vb^2>0$ and $\yb^2>0$. As particular cases, we will discuss the physical content concerning the present time, characterized by $\vb^2=\Omega_m\simeq 0.315$, $\bar{x}^2+\bar{y}^2={ \Omega  }_{ \phi }\simeq 0.685$ and $\omega_m=0$ according to recent observations based on the $\Lambda$CDM model \cite{planck}. We will also analyse the inflationary phase, where we believe there is no matter contribution, corresponding to $\vb^2\to 0$.


\subsection{$Q_1=0$ and $Q_1=\beta$}

For the case $Q_1=0$ there is no interaction between the DM and the ELKO field, thus they evolve independently. The fixed points that follows from the analysis of the system (\ref{xline}) to (\ref{yline}) and satisfy $x'=0$, $v'=0$ and  $y'=0$ are given just by $\xb=1$, $\vb=0$ and $\yb=0$, which does not represent a scaling solution, in the sense that does not admit a mixture of fluids. Besides that our model is valid just for $\yb \neq 0$ and $\vb$ given by (\ref{vbar}).

For the case $Q_1=\beta$, a constant interaction term, we have scaling solutions of the form $\xb\neq 0$, $\vb\neq 0$ and $\yb=0$, which could admit a mixture of the fluids, but the condition $\yb=0$ shows that the potential part of the ELKO field is null according to (\ref{xyv}), leading to $\tilde{V}=0$, and as discussed earlier, this leads to a divergence in the $\lambda$ term, but we have defined it as proportional to the constant $\delta$, so we will discard such kind of fixed point in our analysis. We are interested only in fixed points that satisfy $\xb\neq 0$, $\vb\neq 0$ and $\yb\neq 0$, which are scaling solutions and do not have null potential contributions.

\subsection{$Q_1=\beta x^2$}

Such interaction between DM and ELKO field corresponds to $Q=\beta H \dot{\phi}^2 $, where $\beta$ is a dimensionless parameter. 

From the analysis of the system of equations (\ref{xline2}) and (\ref{vline2}), the fixed points are given by $[\xb,\, \yb,\, \vb]$, with 
\beqq
\bar { x } &=&\sqrt { \frac { \left( 3+3{ \omega  }_{ m }-2\delta  \right) \left( 3+3{ \omega  }_{ \phi  }-2\delta  \right)  }{ 6\beta (\omega _{ \phi  }-\omega _{ m }) }}    \,,\\
\bar{y}&=&\sqrt { \frac { (2\delta - 3\omega_m-3)(2\beta-2\delta+3\omega_\phi+3)  }{ 6\beta  (\omega_\phi -\omega_m )  }  },
\eeqq
and $\bar{v}$ is given by (\ref{vbar}), which is valid for all interactions.

The determinant $\Delta$ and the trace $\tau$ of the matrix of the linearised system (\ref{delta_sys}) to (\ref{d}) are given by
\beqq
\Delta&=&4\delta^2-2\delta(2\beta+6+3(\omega_m+\omega_\phi))+6\beta (1+\omega_m)+9(1+\omega_m\omega_\phi+\omega_m+\omega_\phi)\,,\\
\tau&=&4\delta-2\beta -6 -3(\omega_m+\omega_\phi)\,.
\eeqq 

The stability of the fixed points, namely $\Delta > 0$ and $\tau <0$, is related with the values of $\delta$, $\beta$, $\omega_m$ and $\omega_\phi$. Table \ref{tab_betax2} presents the above conditions plus to reality conditions for the fixed points, namely $\xb^2>0$, $\vb^2>0$ and $\yb^2>0$ for some specific values of equation of state parameter for both DM ($\omega_m$) and ELKO field ($\omega_\phi$).

\begin{table}[t!]
\begin{center}
        \begin{tabular}{|>{\centering\arraybackslash}m{0.25in}|>{\centering\arraybackslash}m{0.5in}|>{\centering\arraybackslash}m{1.25in}|>{\centering\arraybackslash}m{1.25in}|>{\centering\arraybackslash}m{1.25in}|>{\centering\arraybackslash}m{1.25in}|}
            \hline
 & & \multicolumn{4}{c|}{ELKO Spinor} \\ \hline 

& & Vacuum ($\omega_\phi=-1$)& \pbox{20cm}{Dust\\($\omega_\phi=0$)} & Radiation ($\omega_\phi=1/3$) & Ultrarelativistic ($\omega_\phi=1$)\\ \hline

            \multirow{10}{*}{ \begin{turn}{90} Matter \end{turn}} & \begin{turn}{90} \pbox{20cm}{Vacuum \\ ($\omega_m=-1$)}  \end{turn} & ------ & \pbox{20cm}{No stable point}& \pbox{20cm}{No stable point}& \pbox{20cm}{No stable point}
            \\ \cline{2-6}
                                    & \begin{turn}{90} \pbox{20cm}{Dust\\($\omega_m=0$)} \end{turn} & \pbox{20cm}{$\delta\leq 3/2$\\ if $\beta\geq 3/2$\\or \\$\delta <\beta $\\ if $\beta<3/2$}  & ------ & \pbox{20cm}{No stable point} & \pbox{20cm}{No stable point}

            \\ \cline{2-6}
                                    & \begin{turn}{90} \pbox{20cm}{Radiation\\($\omega_m=1/3$)}\end{turn} & \pbox{20cm}{$\delta\leq\beta$ if $\beta\leq 2$\\or\\ $\delta<2$ if $\beta> 2$} & \pbox{20cm}{$\delta\leq 2$\\ if $\beta\geq 1/2$\\or \\$\delta <3/2+\beta $\\ if $\beta< 1/2$} & ------ &\pbox{20cm}{No stable point}
            \\ \cline{2-6}
                                    & \begin{turn}{90} \pbox{20cm}{Ultrarelativistic\\($\omega_m=1$)}\end{turn} & \pbox{20cm}{$\delta\leq 3$ if $\beta\geq 3$\\or \\$\delta <\beta $ if $\beta<3$}& \pbox{20cm}{$\delta\leq 3$ if $\beta\geq 3/2$\\or \\$\delta <\beta+3/2 $\\ if $\beta<3/2$}&\pbox{20cm}{$\delta\leq 3$ if $\beta\geq 1$\\or \\$\delta <\beta+2 $\\ if $\beta<1$}& ------
            \\\hline
          
        \end{tabular}
    \caption{Stability conditions for some equation of state parameters of DM and ELKO field, corresponding to the interaction $Q_1=\beta x^2$}
    \label{tab_betax2}

\end{center}
\end{table}

Now let us analyse the Table \ref{tab_betax2}. We are interested in two different epochs, namely the inflation and the late time acceleration. The first one corresponds to an universe without DM and totally filled with ELKO field. This means $\bar{v}^2={ \Omega  }_{ m }=0$ and $\omega_m=0$, which leads to $\delta={3\over 2}(1+\omega_\phi)$. By replacing into $\xb$ and $\yb$ we obtain $\bar{x}^2=0$ and $\bar{y}^2=1$, which shows that all the contribution should come from the potential part $\tilde{V}$ and also we should have $\dot{\phi}=0$ from (\ref{xyv}), but our interaction term $Q\sim \dot{\phi}$, thus we conclude that such condition can not be applied to the inflation. 

On the other hand, for the present time such variables are given by $\bar{v}^2={ \Omega  }_{ m }=0.315$ and $\bar{x}^2+\bar{y}^2={ \Omega  }_{ \phi }=0.685$ according to the $\Lambda$CDM model, and besides that we must have $\omega_m=0$. From (\ref{vbar}) we conclude that: (i) $\delta\simeq 0.47$ if $\omega_\phi=-1$. From the corresponding cell in the Table \ref{tab_betax2} ($\omega_\phi=-1,\,\,\omega_m=0$), we see that if $\beta \gtrsim 0.47$ the system is stable around the fixed points. Such positive value of $\beta$ corresponds to decay of ELKO field into DM; (ii)  $\delta\simeq 1.84$ if $\omega_\phi=1/3$, but there is no solution for $\beta$ from the Table I in this case; and (iii)  $\delta\simeq 2.53$ if $\omega_\phi=1$, which has no solution for $\beta$ too. We conclude that, for the present time, the only possibility in order to have stable fixed points is $\beta \gtrsim 0.47$, leading to decay of ELKO field into DM for an equation of state parameter $\omega_\phi=-1$, that is, the ELKO field behaviour must be of vacuum type. Notice that other stability conditions are possible if the equation of state parameter of dark matter is of radiation or ultra-relativistic type.

The case $\omega_m=\omega_\phi$ has no physical meaning since both fluids has the same equation of state parameter, thus they are thermodynamically identical.

\subsection{$Q_1=\beta v^2 x^2$}

\begin{table}[t!]
\begin{center}
        \begin{tabular}{|>{\centering\arraybackslash}m{0.25in}|>{\centering\arraybackslash}m{0.5in}|>{\centering\arraybackslash}m{1.25in}|>{\centering\arraybackslash}m{1.25in}|>{\centering\arraybackslash}m{1.25in}|>{\centering\arraybackslash}m{1.25in}|}
            \hline
 & & \multicolumn{4}{c|}{ELKO Spinor} \\ \hline 

& & Vacuum$~ ~$ ($\omega_\phi=-1$)& \pbox{20cm}{Dust\\($\omega_\phi=0$)} & Radiation ($\omega_\phi=1/3$) & Ultrarelativistic ($\omega_\phi=1$)\\ \hline

            \multirow{10}{*}{ \begin{turn}{90} Matter \end{turn}} & \begin{turn}{90} \pbox{20cm}{Vacuum \\ ($\omega_m=-1$)}  \end{turn} & ------ & \pbox{20cm}{No stable point}& \pbox{20cm}{No stable point} &  \pbox{20cm}{No stable point}
            \\ \cline{2-6}
                                    & \begin{turn}{90} \pbox{20cm}{Dust\\($\omega_m=0$)} \end{turn} & \pbox{20cm}{${18\over 2\beta+9}<\delta<{3\over 2}$\\if$\beta>{3\over 2}$}& ------ &  \pbox{20cm}{No stable point}& \pbox{20cm}{No stable point}
            \\ \cline{2-6}
                                    & \begin{turn}{90} \pbox{20cm}{Radiation\\($\omega_m=1/3$)}\end{turn} &  \pbox{20cm}{${16\over \beta+6}<\delta<2$\\if $\beta>2$} &  \pbox{20cm}{${(6\beta+13)\over 4\beta+6}<\delta<2$\\if $\beta>{1\over 2}$}& ------ & \pbox{20cm}{No stable point}
            \\ \cline{2-6}
                                    & \begin{turn}{90} \pbox{20cm}{Ultrarelativistic\\($\omega_m=1$)}\end{turn} & \pbox{20cm}{${36\over \beta+9}<\delta<3$\\if $\beta> 3$}& \pbox{20cm}{${3\over 2}{2\beta+21\over 2\beta+9}<\delta<3$\\if $\beta> {3\over 2}$} & \pbox{20cm}{${2(\beta+5)\over \beta+3}<\delta<3$\\if $\beta> 1$} & ------
            \\\hline
          
        \end{tabular}
    \caption{Stability conditions for some equation of state parameters of DM and ELKO field, corresponding to the interaction $Q_1=\beta v^2 x^2$}
    \label{tab_betav2x2}

\end{center}
\end{table}

For this interaction we have $Q=\frac{1}{3} \kappa^2 H \beta {\rho}_m \dot{\phi}^2$. The fixed points are $[\xb,\,\yb,\,\vb]$, with
\beqq
\bar { x } &=&\sqrt{{3+3\omega_m-2\delta}\over 2\beta}\,,\\
\bar{y}&=&\sqrt { \frac { (2\delta - 3\omega_m-3)(2\beta-3\omega_m+3\omega_\phi)  }{ 6\beta  (\omega_\phi -\omega_m )  }  }\,,
\eeqq
and $\bar{v}$ is given by (\ref{vbar}).

From the linearised matrix one finds:
\beqq
\Delta&=&(4-{8\beta\over 3(\omega_m-\omega_\phi)})\delta^2+{1\over \omega_m-\omega_\phi}[(8+4\omega_m+4\omega_\phi)\beta-6(\omega_m^2-\omega_\phi^2+2\omega_m-2\omega_\phi)]\delta\nonumber\\
&&+{1\over \omega_m-\omega_\phi} [-6\beta(1+\omega_\phi+\omega_m+\omega_\phi\omega_m)-9(\omega_\phi-\omega_m+\omega_\phi^2-\omega_m^2+\omega_m\omega_\phi^2-\omega_\phi\omega_m^2)]\,,\\
\tau&=&(2-{4\beta\over 3(\omega_m-\omega_\phi)})\delta +{1\over \omega_m-\omega_\phi}[2\beta(1+\omega_\phi)-3(\omega_m-\omega_\phi-\omega_\phi^2+\omega_m\omega_\phi)\,.
\eeqq

The Table \ref{tab_betav2x2} shows the stability conditions for some specific values of $\omega_m$ and $\omega_\phi$.

For the inflationary epoch ($\bar{v}^2=0$ and $\omega_m=0$), we have the same condition for $\delta$, namely $\delta={3\over 2}(1+\omega_\phi)$. From the corresponding cell in the Table II it is easy to see that $\omega_\phi=-1$ is the only possible condition of stability, which leads to $\delta=0$, but such value of $\delta$ is not possible from the Table II if $\omega_m=0$. 

For the present time we have $\bar{v}^2={ \Omega  }_{ m }=0.315$. As the previous case, from (\ref{vbar}) we have: (i) $\delta\simeq 0.47$ if $\omega_\phi=-1$, which is possible if $\beta \gtrsim 14.6$ and implies a positive value of $\beta$, consequently the decay of ELKO field into DM. The other cases: (ii)  $\delta\simeq 1.84$ if $\omega_\phi=1/3$, and (iii) $\delta\simeq 2.53$ if $\omega_\phi=1$ does not present stable solutions if $\omega_m=0$.

\subsection{$Q_1=\beta (x^2+y^2)x^2$}

For this interaction we have $Q={1\over 3}\frac{ \kappa^2}{H}\beta {\rho}_\phi \dot{\phi}^2$.

\begin{table}[t!]
\begin{center}
        \begin{tabular}{|>{\centering\arraybackslash}m{0.25in}|>{\centering\arraybackslash}m{0.5in}|>{\centering\arraybackslash}m{1.25in}|>{\centering\arraybackslash}m{1.25in}|>{\centering\arraybackslash}m{1.25in}|>{\centering\arraybackslash}m{1.25in}|}
            \hline
 & & \multicolumn{4}{c|}{ELKO Spinor} \\ \hline 

& & Vacuum$~ ~$ ($\omega_\phi=-1$)& \pbox{20cm}{Dust\\($\omega_\phi=0$)} & Radiation ($\omega_\phi=1/3$) & Ultrarelativistic ($\omega_\phi=1$)\\ \hline

            \multirow{10}{*}{ \begin{turn}{90} Matter \end{turn}} & \begin{turn}{90} \pbox{20cm}{Vacuum \\ ($\omega_m=-1$)}  \end{turn} & ------ & \pbox{20cm}{No stable point}& \pbox{20cm}{No stable point}&  \pbox{20cm}{No stable point}
            \\ \cline{2-6}
                                    & \begin{turn}{90} \pbox{20cm}{Dust\\($\omega_m=0$)} \end{turn} & \pbox{20cm}{$0<\delta <{3\beta\over 3+2\beta}$\\if $\beta>0$}& ------ &  \pbox{20cm}{No stable point}& \pbox{20cm}{No stable point}
            \\ \cline{2-6}
                                    & \begin{turn}{90} \pbox{20cm}{Radiation\\($\omega_m=1/3$)}\end{turn} & \pbox{20cm}{$0<\delta<{2\beta\over 2+\beta}$\\if $\beta>0$} &\pbox{20cm}{${3\over 2}<\delta<{3+8\beta\over 2+4\beta}$\\if $\beta>0$} & ------ & \pbox{20cm}{No stable point}
            \\ \cline{2-6}
                                    & \begin{turn}{90} \pbox{20cm}{Ultrarelativistic\\($\omega_m=1$)}\end{turn} & \pbox{20cm}{$0<\delta<{3\beta\over 3+\beta}$\\if $\beta>0$}& \pbox{20cm}{${3\over 2}<\delta<{9+12\beta\over 6+4\beta}$\\if $\beta>0$} &  \pbox{20cm}{$2<\delta<{2+3\beta\over 1+\beta}$\\if $\beta>0$}& ------
            \\\hline
          
        \end{tabular}
    \caption{Stability conditions for some equation of state parameters of DM and ELKO field, corresponding to the interaction $Q_1=\beta (x^2+y^2) x^2$}.
    \label{tab_betax2x2y2}

\end{center}
\end{table}

The fixed points are $[\xb,\,\yb,\,\vb]$, with
\beqq
\bar {x} &=&\sqrt{2\delta-3-3\omega_\phi\over 2\beta}\,,\\
\bar{y}&=&\sqrt { \frac { 2\delta(2\beta+3\omega_m - 3\omega_m)-6\beta(1+\omega_m)+9(\omega_\phi^2-\omega_m\omega_\phi-\omega_m+\omega_\phi)  }{ 6\beta  (\omega_\phi -\omega_m )  }  }\,,
\eeqq
and $\bar{v}$ given by (\ref{vbar}).

From the linearised matrix one finds:
\beqq
\Delta&=&(4+{8\beta\over 3(\omega_m-\omega_\phi)})\delta^2-{1\over \omega_m-\omega_\phi}[8\beta(1+\omega_m)+6(\omega_m^2-\omega_\phi^2+2\omega_m-2\omega_\phi)]\delta\nonumber\\
&&+{1\over \omega_m-\omega_\phi} [6\beta(1+\omega_m^2+2\omega_m)+27(\omega_\phi\omega_m^2+\omega_m^2-\omega_\phi^2\omega_m+\omega_m-\omega_\phi^2-\omega_\phi^2-\omega_\phi)]\,,\\
\tau&=&(2+{4\beta\over 3(\omega_m-\omega_\phi)})\delta +{1\over \omega_m-\omega_\phi}[-2\beta(1+\omega_m)-3(\omega_m-\omega_\phi+\omega_m^2-\omega_m\omega_\phi)]\,.
\eeqq
Table \ref{tab_betax2x2y2} shows the stability conditions for some types of $\omega_m$ and $\omega_\phi$.

For the inflationary epoch ($\vb^2=0$ and $\omega_m=0$) we have $\bar{x}^2=0$ and $\bar{y}^2=1$, which shows that the contribution comes only from the potential part. The kinetic part is null, but the interaction is proportional to $\dot{\phi}^2$, thus this interaction does not applies to inflation. 

Considering the present time ($\bar{v}^2=0.315$) we have: (i) $\delta\simeq 0.47$ if $\omega_\phi=-1$. We have verified that for $\beta \gtrsim 0.69$ we have $\xb^2>0$ and $\yb^2>0$, which represents stable fixed points. The cases (ii) $\delta\simeq 1.84$ if $\omega_\phi=1/3$, and (iii) $\delta\simeq 2.53$ if $\omega_\phi=1$ does not present stable fixed points for $\omega_m=0$.

\subsection{$Q_1=\beta (v^2 - y^2)$}

The corresponding interaction is $Q=2\beta H (\rho_m-\rho_\phi+{1\over 2}\dot{\phi}^2)$. The fixed points are $[\xb,\,\yb,\,\vb]$, with
\beqq
\xb&=&\sqrt{\frac{4\delta^2+2\delta(4\beta-6-3(\omega_m+\omega_\phi))-6\beta(2+\omega_m+\omega_\phi)+9(1+\omega_m+\omega_\phi+\omega_m\omega_\phi)}{6\beta(\omega_\phi-\omega_m)}}\,,\\
\yb&=& \sqrt{\frac{(2\delta-3-3\omega_\phi)(2\delta+2\beta-3-3\omega_m)}{6\beta(\omega_m-\omega_\phi)}}\,,
\eeqq
and $\bar{v}$ given by (\ref{vbar}).

The determinant and trace are given by:
\beqq
\Delta &=& 4\delta^2+2\delta(2\beta-6-3(\omega_\phi+\omega_m))-6\beta(1+\omega_\phi)+9(1+\omega_m+\omega_\phi+\omega_m\omega_\phi)\,,\\
\tau &=& 4\delta -6 +2\beta -3(\omega_m+\omega_\phi)\,.
\eeqq

Table \ref{tab_betav2my2} shows the stability conditions for some specific values of $\omega_m$ and $\omega_\phi$.

\begin{table}[t!]
\begin{center}
        \begin{tabular}{|>{\centering\arraybackslash}m{0.25in}|>{\centering\arraybackslash}m{0.5in}|>{\centering\arraybackslash}m{1.25in}|>{\centering\arraybackslash}m{1.25in}|>{\centering\arraybackslash}m{1.25in}|>{\centering\arraybackslash}m{1.25in}|}
            \hline
 & & \multicolumn{4}{c|}{ELKO Spinor} \\ \hline 

& & Vacuum$~ ~$ ($\omega_\phi=-1$)& \pbox{20cm}{Dust\\($\omega_\phi=0$)} & Radiation ($\omega_\phi=1/3$) & Ultrarelativistic ($\omega_\phi=1$)\\ \hline

            \multirow{10}{*}{ \begin{turn}{90} Matter \end{turn}} & \begin{turn}{90} \pbox{20cm}{Vacuum \\ ($\omega_m=-1$)}  \end{turn} & ------ &  \pbox{20cm}{$\delta_1<\delta<{-\beta}$\\if $-{3\over 2}\leq\beta<0$\\or\\$\delta_1<\delta<{3\over 2}$\\if $\beta<-{3\over 2}$ } & \pbox{20cm}{$\delta_2<\delta<{-\beta}$\\if $-2\leq\beta<0$\\or\\$\delta_2<\delta<2$\\if $\beta<-2$ } & \pbox{20cm}{$\delta_3<\delta<{-\beta}$\\if $-3\leq\beta<0$\\or\\$\delta_3<\delta<3$\\if $\beta<-3$ } 
            \\ \cline{2-6}
                                    & \begin{turn}{90} \pbox{20cm}{Dust\\($\omega_m=0$)} \end{turn} &\pbox{20cm}{No stable point}& ------ &  \pbox{20cm}{$\delta_4<\delta<{3\over 2}-\beta$\\if $-{1\over 2}\leq\beta<0$\\or\\$\delta_4<\delta<2$\\if $\beta<-{1\over 2}$ }& \pbox{20cm}{$\delta_5<\delta<{3\over 2}-\beta$\\if $-{3\over 2}\leq\beta<0$\\or\\$\delta_5<\delta<3$\\if $\beta<-{3\over 2}$ }
            \\ \cline{2-6}
                                    & \begin{turn}{90} \pbox{20cm}{Radiation\\($\omega_m=1/3$)}\end{turn} &  \pbox{20cm}{No stable point}& \pbox{20cm}{No stable point}& ------ &  \pbox{20cm}{$\delta_6<\delta<{2}-\beta$\\if $-{1}\leq\beta<0$\\or\\$\delta_6<\delta<3$\\if $\beta<-{1}$ }
            \\ \cline{2-6}
                                    & \begin{turn}{90} \pbox{20cm}{Ultrarelativistic\\($\omega_m=1/3$)}\end{turn} &  \pbox{20cm}{No stable point}& \pbox{20cm}{No stable point}& \pbox{20cm}{No stable point} &  ------
            \\\hline
          
        \end{tabular}
    \caption{Stability conditions for some equation of state parameters of DM and ELKO field, corresponding to the interaction $Q_1=\beta (v^2-y^2)$. We have defined the following parameters: $\delta_1={3\over 4}-\beta-{1\over 4}\sqrt{9+16\beta^2}$; $\delta_2=1-\beta-\sqrt{1+\beta^2}$; $\delta_3={3\over 2}-\beta-{1\over 2}\sqrt{9+4\beta^2}$; $\delta_4={7\over 4}-\beta-{1\over 4}\sqrt{1+16\beta^2}$; $\delta_5={9\over 4}-\beta-{1\over 4}\sqrt{9+16\beta^2}$; $\delta_6={5\over 2}-\beta-{1\over 2}\sqrt{1+4\beta^2}$; }.
    \label{tab_betav2my2}

\end{center}
\end{table}

For the inflationary epoch ($\bar{v}^2=0$ and $\omega_m=0$), we have the same condition for $\delta$, namely $\delta={3\over 2}(1+\omega_\phi)$. Contrary to previous cases, we see that $\omega_\phi=-1$ is not a stable solution. For $\omega_\phi=1/3$ we have $\delta = 2$, which is a stable solution corresponding to $\beta = -1/2$, representing the decay of DM into ELKO field. But we have inferred $\vb^2=0$, thus there is no matter to decay at the inflation epoch. For $\omega_\phi = 1$ we have $\delta = 3$, but the stability condition requires $\delta < 3$ from the corresponding cell in Table IV. We conclude that such interaction does not present stable points for the inflation.

For the present time we have $\bar{v}^2=0.315$ and $\omega_m=0$. As the previous case, we have: (i) $\delta\simeq 0.47$ if $\omega_\phi=-1$, which has no stable solution; (ii)  $\delta\simeq 1.84$ if $\omega_\phi=1/3$, and it is easy to see from Table IV that such value of $\delta$ is possible for a negative value of $\beta$. For instance, if $\beta=-1/2$ the condition for $\delta$ is $1.69\lesssim \delta \lesssim 2$; and (iii) $\delta\simeq 2.53$ if $\omega_\phi=1$, and for this condition we also have stable solution for a negative $\beta$, as can be seen from Table IV. If $\beta = -3/2$ for instance, the condition on $\delta$ is $2.07\lesssim \delta \lesssim 3.0$, which includes $\delta\simeq 2.53$. Thus, contrary to the previous cases, if the ELKO equation of state parameter is of radiation or ultrarelativistic type, the system presents stable solutions if $\beta$ is negative, which corresponds to the decay of DM into ELKO field. The present acceleration of the universe can be understood in this model as the decay of dark matter into ELKO particles. This is a very interesting scenery which also alleviates the cosmic coincidence problem.

\section{Concluding remarks}

In this work we have developed a new approach to study the stability of a system composed by an ELKO field interacting with DM, which could give some contribution in order to alleviate the cosmic coincidence problem. Since that recent works \cite{WEI,basak,sadja,saj} have not found stable points for such system for different dynamic variables and interactions terms, we are led to believe (without demonstration) that the system ELKO-DM does not allow stable points. Based on these results, we have supposed an additional constraint to the dynamical system, namely that the potential of the ELKO field is related to a constant parameter $\delta$, then we have analysed the stability conditions for such new system. We have also assumed that both the ELKO field as the dark matter energy density are related to the pressure by equations of state parameters $\omega_\phi$ and $\omega_m$, respectively. We have found different stability conditions relating the parameter $\delta$ and the interaction parameter $\beta$, which states if the decay is from DM to ELKO ($\beta <0$) or from ELKO to DM ($\beta >0$). Different values of $\omega_\phi$ and $\omega_m$ corresponding to vacuum, dust, radiation and ultra-relativistic equation of state parameter are presented in Tables I to IV for different interaction terms, showing the conditions for the existence of stable points. 

For the first three tables, corresponding to the interactions $Q_1=\beta x^2$, $Q_1=\beta v^2 x^2$ and $Q_1=\beta(x^2+y^2)x^2$, the conditions for stable fixed points in order to satisfy the present stage of acceleration (with $\omega_m=0$) are given by positive $\beta$ and $\delta$ parameters and also require an equation of state parameter for the ELKO field of vacuum type ($\omega_\phi=-1$).  Positive values of $\beta$ means the decay of ELKO field into DM particles. Such conditions could alleviate the cosmological coincidence problem. The inflationary phase cannot be driven for these interactions if we set $\omega_m=0$ and $\vb^2=0$. Other possibilities are allowed if the equation of state parameters of DM and ELKO field are of radiation or ultra-relativistic type. Another interesting aspect that follows from Tables I to III is that there is no stable fixed points if $\omega_\phi > \omega_m$.

For the last interaction, namely $Q_1=\beta(v^2-y^2)$, we have the opposite. The conditions for stable fixed points in order to satisfy the present stage of acceleration are given by negative $\beta$ and positive $\delta$ parameters. The equation of state parameter for the ELKO field must be of radiation ($\omega_\phi=1/3$) or ultra-relativistic ($\omega_\phi=1$) type. The case $\omega_\phi=-1$ does not present stable fixed points. Negative values of $\beta$ means the decay of DM particles into ELKO field. Such conditions also could alleviate the cosmological coincidence problem, and the equation of state parameter of ELKO field is not of exotic type. The inflationary phase cannot be driven for this interaction too. Other possibilities are allowed if the equation of state parameters of DM and ELKO field are of radiation or ultra-relativistic type. Another interesting aspect that follows from Table IV is that, contrary to the previous cases, there is no stable fixed points if $\omega_\phi < \omega_m$.

In such analysis the interaction $Q$ must be proportional to the variable $x$, otherwise the relations among the parameters must be very restrictive. But such condition is not so strong, since that the variable $x$ is proportional to $\dot{\phi}$, which characterizes a time dependence of the field $\phi$, and it is reasonable for an interacting theory. Notice that all the interactions studied are proportional to $\dot{\phi}^2$. For all the interactions analysed here there are conditions of stability in order to alleviate the cosmic coincidence problem. Such kinds of interactions and conditions on the parameters $\beta$ and $\delta$ open possibilities for future searches concerning the interaction between DM and ELKO field for specific potentials satisfying the conditions presented in the tables. 

Finally, a general theory that includes torsion in ELKO cosmology is of great interest, and in fact it may be the responsible for the late time acceleration of the universe \cite{FABBRI1}. From a dynamical system approach the system of dynamical equation should be rewritten for new variables, certainly with some interesting consequences, but this is left for future works.



\begin{acknowledgements}
SHP is grateful to CNPq - Conselho Nacional de Desenvolvimento Cient\'ifico e Tecnol\'ogico, Brazilian research agency, for the financial support, process number 477872/2010-7. JFJ is grateful to Unesp - C\^ampus Guaratinguet\'{a} and to SHP for hospitality and facilities. We would like to thank the referee for the suggestions and useful comments, especially in which concerns the general treatment including torsion.
\end{acknowledgements}


\end{document}